# Machine Learning Electronic Hamiltonian with Long-Range Effects and External Fields

Xueheng Kuang[1], Ruoyu Wang[1], Kwangrae Kim[1], Rongjing Guo[1], Yiyang Fu[1], and Yuanyue Liu[1,*]

Texas Materials Institute and Department of Mechanical Engineering, The University of Texas at Austin, Austin, Texas 78712, USA

[*]E-mail: Yuanyue.liu@austin.utexas.edu

**Abstract**

Conventional machine-learning electronic Hamiltonian (MLEH) models rely on a local-environment approximation and can miss the effects of long-range (LR) interactions and external fields. Here, we introduce MLEH-LR, a framework that incorporates these effects into Hamiltonian prediction. Benchmarks across materials with different polarizations and dimensionalities demonstrate substantial improvements in the accuracy of the predicted Hamiltonians and resulting electronic structures. MLEH-LR enables accurate, large-scale, high-throughput electronic-structure calculations, as demonstrated for zigzag boron nitride ribbons under transverse electric fields. These calculations allow us to formulate a quantitative relation describing the joint width and field dependence of the band gap and to reveal the electronic mechanisms underlying its field response. Our work provides an efficient method and tool for studying electronic structure in materials where long-range electrostatics and external fields play an important role.

## 1 Introduction

Machine-learning electronic Hamiltonian (MLEH) models provide a powerful approach to studying the electronic properties of molecules and materials[1–7]. By predicting Hamiltonian matrix elements from atomic structures, they enable calculations of electronic bands, orbitals, and related properties without repeating costly electronic-structure calculations. Representative models, including DeepH and DeepH-E3[3,4], HamGNN[5], DeePTB-E3[6], and MACE-H[7], preserve rotational covariance through local coordinate frames or equivariant representations, enabling efficient electronic-structure calculations across diverse atomic configurations.

However, conventional MLEH models often rely on a local-environment approximation. Their Hamiltonian matrix elements between localized orbitals are predicted from finite atomic neighborhoods, which can miss long-range electrostatic effects from distant charges or polarization. Such effects are particularly important in electrically polar systems[8,9]. Moreover, even at fixed atomic geometry, an applied electric field changes the Hamiltonian through both the external potential and the induced electronic density. Capturing these effects requires information beyond local geometry.

Similar limitations also arise in machine-learning interatomic potentials and force fields, and several solutions have been developed. For example, charge-equilibration models capture nonlocal charge redistribution by determining atomic charges collectively across the system[10]. MACE-POLAR-1[11] further describes polarization through learned atomic multipoles that are updated in response to the electrostatic potential, while enforcing a global charge constraint. These approaches primarily target energies and forces.

Long-range treatments are less developed for machine-learning electronic Hamiltonians. Recently, HamGNN-LR was introduced to combine reciprocal-space Ewald attention with an analytic long-range Coulomb correction[12]. Although its source predictions include nonlocal information, the electrostatic sources are represented by atomic monopoles while neglecting higher-order moments, and it does not explicitly update these sources in response to their computed potential. Furthermore, it lacks an explicit applied-field input, so it cannot directly describe field-induced changes to the Hamiltonian at fixed atomic geometry.

Other learned-Hamiltonian approaches have addressed the response to applied electric fields. For example, Bhat et al.[13] combine a learned density-matrix-dependent Hamiltonian with a prescribed time-dependent potential to simulate electron dynamics. DeePTB-NEGF couples learned Hamiltonians with NEGF–Poisson calculations to describe gate and bias effects[14]. However, these approaches still require time propagation or electronic self-consistency to obtain the induced electronic response. Neither directly predicts the converged static Hamiltonian from atomic geometry and a prescribed field. This motivates a framework that directly predicts field-dependent static Hamiltonians, including the induced electronic redistribution, for efficient large-scale exploration of structures and fields.

Here, we introduce a framework for machine-learning electronic Hamiltonians with both long-range effects and external fields (MLEH-LR). Learned atomic monopoles, dipoles, and quadrupoles define a smooth charge density, and solving the Poisson equation provides the long-range potential. This potential feeds back into learned nonlinear multipole updates, explicitly connecting the long-range electrostatic environment to changes in the source distribution. To incorporate an external field, we include its prescribed potential in both the learned multipole updates and the prediction of Hamiltonian matrix elements. This allows the model to account for both the induced electronic redistribution and the direct contribution of the external potential. We compare different implementations of this framework by benchmarking across one-dimensional h-BN ribbons, two-dimensional ZnO slabs, and three-dimensional GaN/AlN and CdSe/ZnS superlattices. We find substantial improvements in Hamiltonian and band predictions over the conventional local MLEH model. Applying MLEH-LR to h-BN ribbons under transverse electric fields, we exploit its transferability across fields and ribbon widths to map electronic structure over a broad field–width space without repeating costly self-consistent calculations for each combination. This capability allows us to establish a quantitative relation for the joint width and field dependence of the band gap and reveal the microscopic origin of the band-gap maximum.

## 2 Method

### 2.1 Architecture Overview

Our implementation is based on an existing short-range machine-learning electronic Hamiltonian (MLEH) model. In this work, we implement our MLEH-LR framework using MACE-H, but it could also be implemented in other MLEH models, such as DeepH and HamGNN. As shown in Figure 1, we first use MACE[15] to convert atomic positions $\boldsymbol{R}_i$ and species $Z_i$ into node features $\boldsymbol{h}_i^{(t)}$, where $t$ labels the message-passing layer and $i$ the atomic site. The short-range parts of $H$ are predicted from local edge features ($\boldsymbol{e}_{ij}$) derived from the node features as implemented in MACE-H. Meanwhile, a long-range branch converts the node features into an electrostatic potential $\phi_i$ on each atom. These potentials, together with the applied potential from the external field, contribute to the prediction of the final Hamiltonian $H$ in different ways.

We use three methods to incorporate the electrostatic potentials: (1) feature injection (FI), which incorporates refined multipoles into node features and atomic-site potentials into edge features; (2) operator correction (OC), which adds an explicit orbital-overlap-weighted potential correction to a local Hamiltonian; and (3) two-stage potential conditioning (TPC), which tests whether learning electrostatic potentials separately, rather than jointly with the Hamiltonian, can improve Hamiltonian prediction. In TPC, a potential predictor is first trained on atomic-site Hartree potentials, followed by a separate Hamiltonian model conditioned on these potentials.

Below, we describe the key components of the MLEH-LR models.

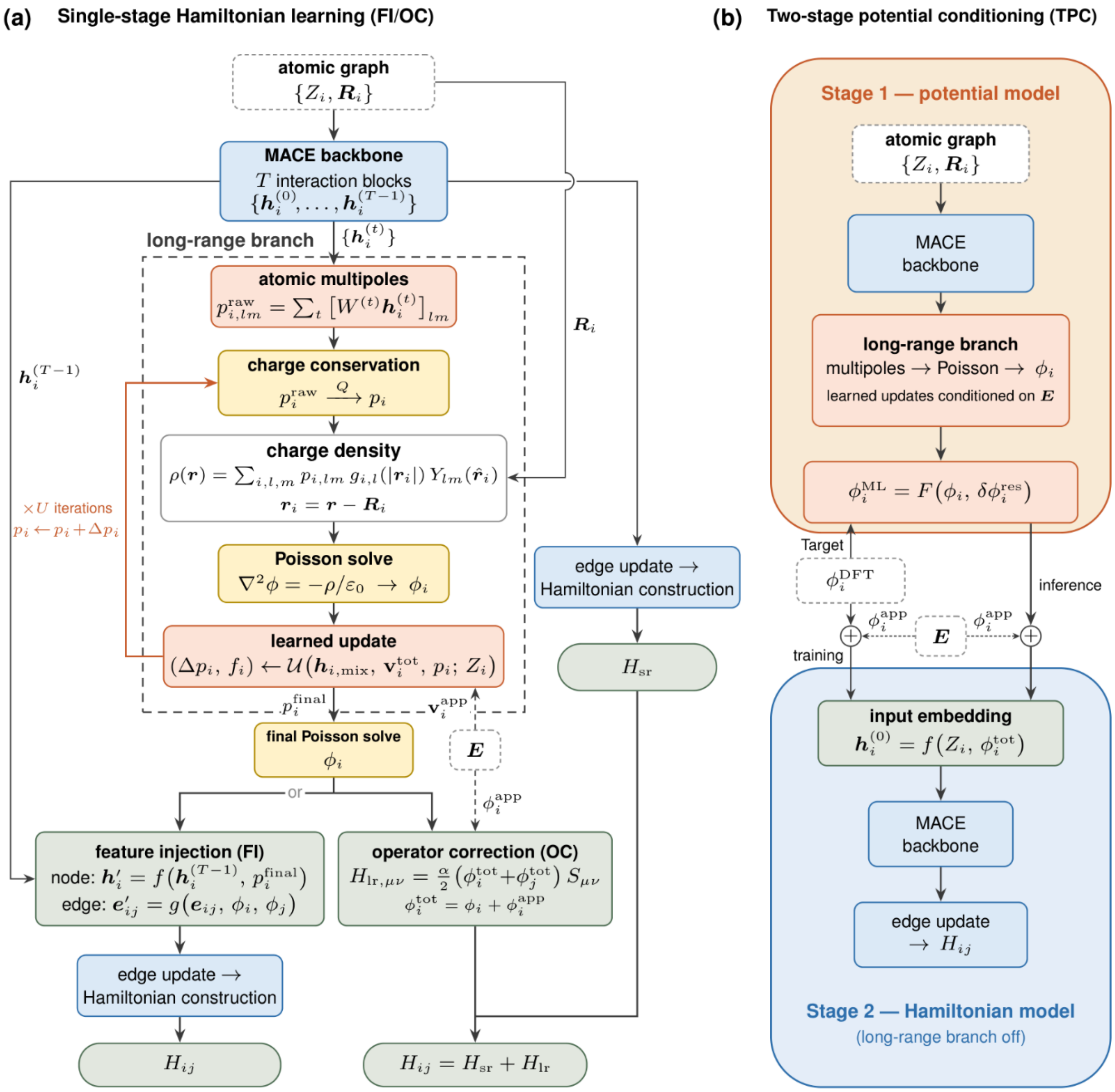

**Figure 1.** Architecture of the machine-learning electronic Hamiltonian model with long-range effects and external fields (MLEH-LR). (a) Single-stage feature injection (FI) and operator correction (OC) models. They start from equivariant maps of the local node features, followed by a long-range branch that produces learned intermediate and final potentials. FI supplies the final multipoles to node features and atomic-site potentials to edge features while OC adds the overlap-weighted correction of Eq. (8) to a local Hamiltonian from MACE-H. The dashed $E$ box represents the prescribed external field. Its upward arrow supplies the applied-potential descriptors to each update, and its downward arrow supplies the applied site potential to the total-potential operator. (b) Two-stage potential conditioning (TPC) model. Stage 1 predicts atomic-site Hartree potentials using DFT supervision with the long-range branch in (a). Stage 2 is trained with the DFT Hartree potential plus the known applied potential and uses the predicted Hartree potential from Stage 1 plus the same applied term at inference.

## 2.2 Long-range branch

The local node features generated by MACE-H are first converted to atomic multipole coefficients, $p_{i,lm}^{raw} = \sum_t \left[W^{(t)}\boldsymbol{h}_i^{(t)}\right]_{lm}$ (1), where $W^{(t)}$ are equivariant linear maps[16]. To ensure charge conservation (the sum of monopoles equals the predefined total charge $Q$ of the system), we renormalize the monopoles over the atoms using learned Fukui-type weights $f_i$:[11]

$$p_{i,00} \leftarrow p_{i,00}^{raw} + \frac{f_i}{\sum_j f_j}\left(Q - \sum_j p_{j,00}^{raw}\right), \qquad (2)$$

where the sum runs over the atoms in one structure. The multipoles then define a smooth charge density:

$$\rho(\boldsymbol{r}) = \sum_i \sum_{l=0}^{l_{\max}} \sum_{m=-l}^{l} p_{i,lm}\, g_{i,l}(|\boldsymbol{r} - \boldsymbol{R}_i|) Y_{lm}\left(\widehat{\boldsymbol{r} - \boldsymbol{R}_i}\right), \qquad (3)$$

Here $g_{i,l}(|\boldsymbol{r} - \boldsymbol{R}_i|)$ are radial functions, and $Y_{lm}$ are real orthonormal spherical harmonics. $\boldsymbol{R}_i$ is the position of atom $i$, and $\widehat{\boldsymbol{r} - \boldsymbol{R}_i}$ denotes the direction from that atom. We consider two types of radial functions: a Gaussian radial function with a fixed width shared by all atoms, and a learned, atom-dependent radial function constructed using a neural network. Their construction is detailed in SI Section S1. In principle, other radial functions, including the fully learned radial function in EdenGNN[17], numerical auxiliary functions in SALTED[18], and Slater-type multipole functions[19], could also be incorporated into our framework.

Once the charge density is determined, the electrostatic potential is obtained by solving the Poisson equation,

$$-\varepsilon_0 \nabla^2 \phi(\boldsymbol{r}) = \rho(\boldsymbol{r}) \quad (4)$$

with boundary conditions appropriate to the target system. The reciprocal-space solution and boundary treatment used here are given in SI Section S2.

The electrostatic potential at each atom is then fed back to update its multipoles, incorporating the LR effects of distant environments:

$$(\Delta p_i, f_i) = \mathcal{U}(\boldsymbol{h}_{i,mix}, \boldsymbol{v}_i, p_{i,lm}; Z_i), \qquad p_i \leftarrow p_i + \Delta p_i, \qquad (5)$$

Here $p_i$ collects the multipole coefficients at site $i$. The update network $\mathcal{U}$ uses equivariant tensor products and nonlinear scalar coefficients. $\boldsymbol{h}_{i,mix}$ is a learned equivariant linear combination of node features from all backbone layers. The potential descriptor $\boldsymbol{v}_i$ contains potential information from $\phi_i$ as node features.

*External potential.*

An external field $\mathbf{E}$ can be included through its atomic-site potential values and the corresponding update descriptors. These enter as

$$\boldsymbol{v}_i^{tot} = \boldsymbol{v}_i^{\rho} + \boldsymbol{v}_i^{app}, \qquad \phi_i^{tot} = \phi_i + \phi_i^{app}, \qquad (6)$$

where $\phi_i^{app} = \phi_{app}(\boldsymbol{R}_i)$. Here, $\boldsymbol{v}_i^{\rho}$ is the descriptor of the Poisson potential $\phi_i$ generated by the learned charge density, and $\boldsymbol{v}_i^{app}$ is that of the applied external potential. The total descriptor conditions the learned multipole updates in Eq. (5), and the total scalar potential enters the Hamiltonian coupling. In the finite-field benchmark, the external potential is the linear part of a sawtooth potential, $\phi_{app}(\mathbf{r}) = -\mathbf{E} \cdot (\mathbf{r} - \bar{\mathbf{r}})$, referenced to the mean atomic position $\bar{\mathbf{r}}$. Implementation details of the multipole updates with and without an external field are given in SI Section S3.

2.3 **Hamiltonian prediction with long-range effects**

After the last learned update, the long-range branch supplies refined multipoles and the scalar atomic-site potential $\phi_i$. Depending on how this final electrostatic information is used, we consider three MLEH-LR models:

(1) Feature injection (FI). We use the refined multipoles to correct the node features through an equivariant map, $\boldsymbol{h}_i' = f\left(\boldsymbol{h}_i^{(T-1)}, p_i^{final}\right)$. In addition, atomic-site potential differences $\phi_j - \phi_i$ and averages $\frac{1}{2}(\phi_i + \phi_j)$ modulate the existing edge features through a residual map, $\boldsymbol{e}_{ij}' = g(\boldsymbol{e}_{ij}, \phi_i, \phi_j)$. The modified node and edge features then enter a final MACE-H edge-update block to predict the Hamiltonian.

(2) Operator correction (OC). This route instead adds an explicit long-range Hamiltonian correction $H_{lr}$ to a short-range Hamiltonian $H_{sr}$ provided by MACE-H,

$$H = H_{sr} + H_{lr}, \qquad (7)$$

using the overlap-weighted form

$$H_{lr,\mu\nu} = \frac{\alpha}{2}(\phi_i + \phi_j)S_{\mu\nu} \qquad (\mu \in i,\ \nu \in j), \qquad (8)$$

Here $\mu$ and $\nu$ label atomic orbitals centered at sites $i$ and $j$, and $S_{\mu\nu} = \langle \mu \mid \nu \rangle$ is their overlap matrix element, and $\alpha$ is a fixed or learned global coefficient. Under an external potential, the site potential is replaced by the total potential from Eq. (6). This construction follows the form of the self-consistent-charge density-functional tight-binding (SCC-DFTB) correction.[20]

(3) Two-stage potential conditioning (TPC). TPC separates potential prediction (Stage 1) from Hamiltonian prediction (Stage 2). In Stage 1, a model is trained to predict the DFT Hartree potential at each atom. It combines the Poisson-derived site potential $\phi_i$ with a learned local residual:

$$\phi_i^{\mathrm{ML}} = \phi_i + sigmoid\left(\eta(\boldsymbol{d}_i)\right) \delta\phi_i^{\mathrm{res}}(\boldsymbol{d}_i), \qquad (9)$$

Here, $\phi_i^{\mathrm{ML}}$ is the machine-learned atomic-site potential, $\delta\phi_i^{\mathrm{res}}$ is a learned local residual potential term, and $\eta$ is its gate network. The optional local residual term compensates for errors in the long-range Hartree potential from the LR branch, in which higher-order multipoles beyond the quadrupole are neglected in the charge-density construction. Then $\phi_i^{\mathrm{ML}}$ is used in Stage 2, which adds the total potential $\phi_i^{\mathrm{tot}}$ to the node features of MACE-H:

$$\boldsymbol{h}_i^{(0)} = W_Z\, e_{Z_i} + W_\phi\, \phi_i^{\mathrm{tot}}, \qquad (10)$$

where $\boldsymbol{h}_i^{(0)}$ is the initial node feature, $e_{Z_i}$ is the one-hot species vector, and $W_Z$ and $W_\phi$ are learned weights. In Stage 2, we first train the MLEH-LR model with the scalar potential $\phi_i^{\mathrm{tot}} = \phi_i^{\mathrm{DFT}} + \phi_i^{\mathrm{app}}$ in the node features of MACE-H. Once we obtain $\phi_i^{\mathrm{ML}}$ from Stage 1, we can predict the Hamiltonian using the trained MLEH-LR model supplied with $\phi_i^{ML} + \phi_i^{\mathrm{app}}$ . Further implementation details of FI, OC, and TPC are given in SI Section S4.

## 3 Results

### 3.1 Hamiltonian accuracy

We evaluate four polar systems spanning three physical dimensionalities as shown in Figure 2: hexagonal BN ribbons (1D), ZnO slabs (2D), and GaN/AlN and CdSe/ZnS superlattices (3D). Training and validation datasets used randomly perturbed structures at a range of widths or thicknesses, while the test dataset used the unperturbed structures over the same width/thickness ranges. Reference Hamiltonians are obtained from self-consistent, spin-unpolarized PBE calculations with OpenMX and localized pseudo-atomic orbitals.[21] Dataset construction, DFT settings, and training parameters are given in SI Section S5.

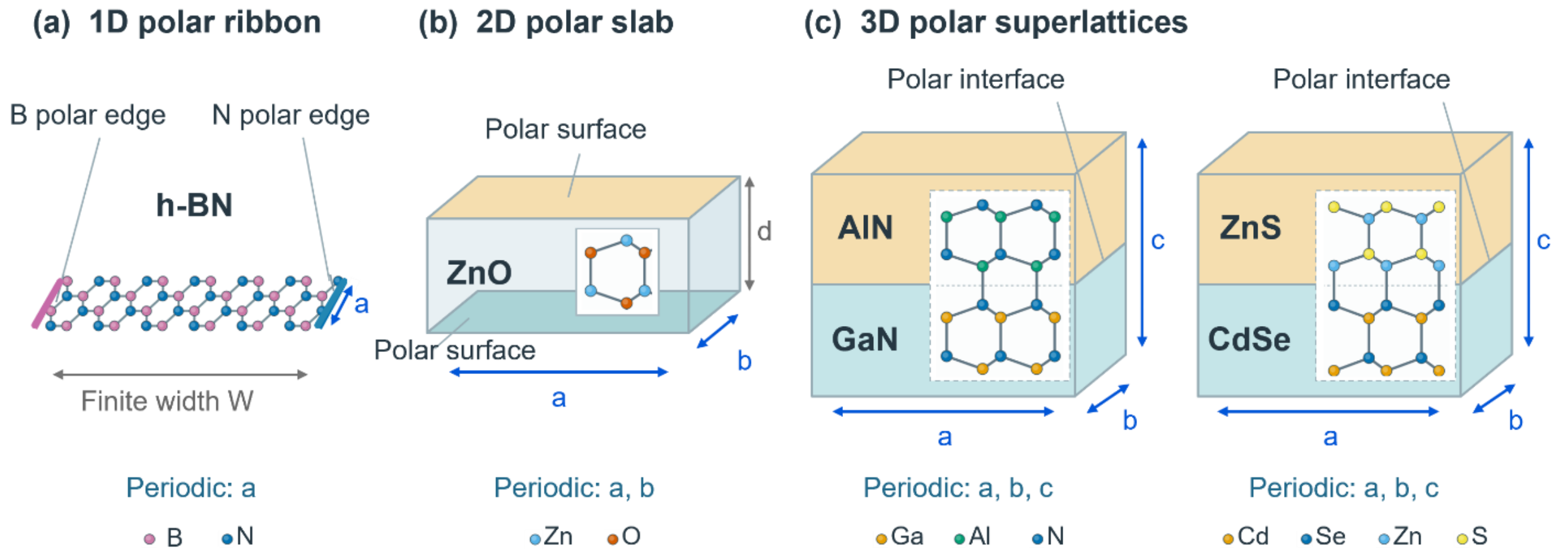

**Figure 2.** Representative test structures spanning three physical dimensionalities: (a) 1D h-BN ribbon with polar edges, (b) 2D ZnO slab with polar surfaces, and (c) 3D GaN/AlN and CdSe/ZnS superlattices with polar interfaces. Blue arrows indicate the periodic lattice directions: a in (a), a and b in (b), and a, b, and c in (c). Gray arrows mark the finite ribbon width W in (a) and slab thickness d in (b).

**Table 1.** Test Hamiltonian MAEs (meV) for the short-range MACE-H model and MLEH-LR models. FI, OC, and TPC denote feature injection, operator correction, and two-stage potential conditioning models with the same learned radial-function backend. Separate columns give $U = 2$ and $U = 1$ learned multipole updates; for TPC, $U$ refers to Stage 1. Bold identifies the lowest long-range error in each row.

| System | Short range | FI | | OC | | TPC | |
|---|---|---|---|---|---|---|---|
| | MACE-H | $U = 2$ | $U = 1$ | $U = 2$ | $U = 1$ | $U = 2$ | $U = 1$ |
| GaN/AlN | 3.112 | 1.378 | 1.539 | 0.870 | 0.810 | **0.759** | 2.186 |
| h-BN | 2.635 | 0.501 | 1.050 | 0.614 | **0.374** | 0.806 | 1.172 |
| ZnO | 3.369 | 2.196 | 2.228 | **1.617** | 2.363 | 2.266 | 3.087 |
| CdSe/ZnS | 1.329 | 0.402 | 0.393 | 0.562 | 0.426 | **0.329** | 0.556 |

Table 1 compares the Hamiltonian errors of the short-range MACE-H with those of the FI, OC, and TPC long-range models introduced in the previous section. These long-range models use the learned radial function. All models use three message-passing blocks. All the long-range models have lower errors than the short-range model, and the best model (highlighted in bold) reduces the error by factors of $2.1–7.0$ across the four systems. The preferred iteration depth $U$ depends on the route. $U = 2$ improves FI in three of the four systems, whereas OC favors $U = 1$ except for ZnO. TPC favors $U = 2$ in all four systems. In the subsequent zero-field analyses, we use the $U$ that gives the lowest mean test Hamiltonian MAE in Table 1 for each method and material system.

TPC introduces an additional dependence on the accuracy of its intermediate potential. For example, replacing the learned radial function in Stage 1 with the fixed radial function increases the potential error on the two 3D superlattices, which raises the corresponding Hamiltonian MAEs from $0.76$ to $37.5$ meV on GaN/AlN and from $0.33$ to $4.2$ meV on CdSe/ZnS. Among the models in Table 1, OC has the lowest cross-system mean Hamiltonian MAE: 0.81 meV, compared with 1.04 meV for TPC and 1.12 meV for FI. Across the four systems, the operator reduces the error by a factor of about 3.7 relative to the local three-block baseline, compared with 2.8–3.0 for the other long-range routes. The performance of OC also motivates its use for finite-field prediction in Section 3.3. Results for FI, OC, and TPC with the fixed radial function are given in SI Section S6.

### 3.2 Size dependence and electronic structure

Figure 3 shows the Hamiltonian errors for individual structures. The accuracy of the short-range MACE-H model generally decreases as the polar structure grows. The long-range models have lower errors throughout the tested size ranges, indicating that the explicit electrostatic representation reduces the size-dependent degradation of the local model.

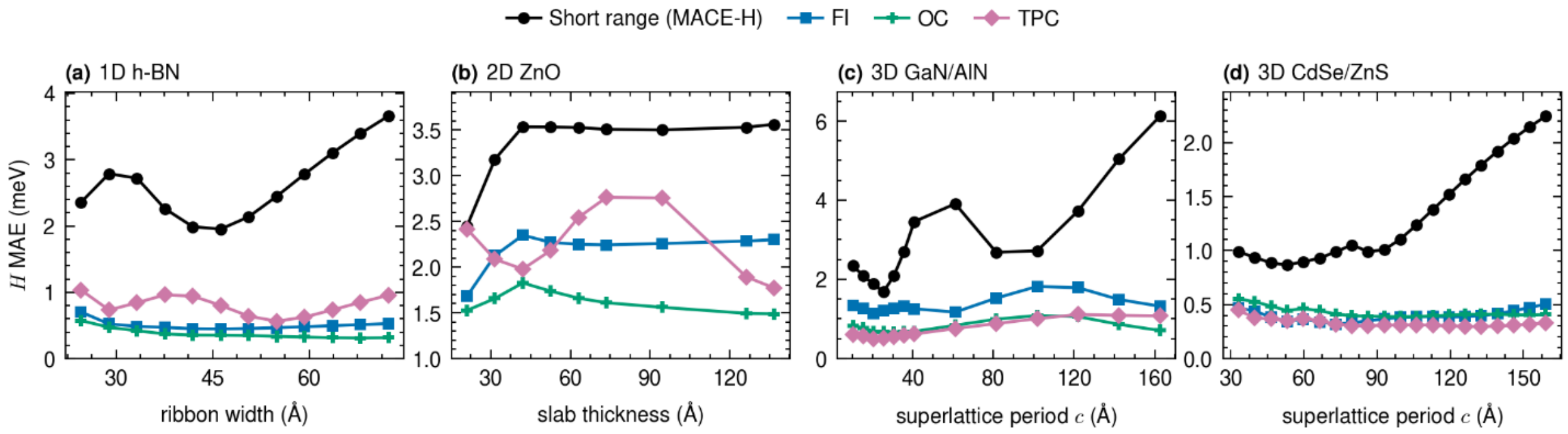

**Figure 3.** Test Hamiltonian MAE versus size for (a) 1D h-BN, (b) 2D ZnO, (c) 3D GaN/AlN and (d) 3D CdSe/ZnS. The short-range model is compared with FI, OC, and TPC from Table 1.

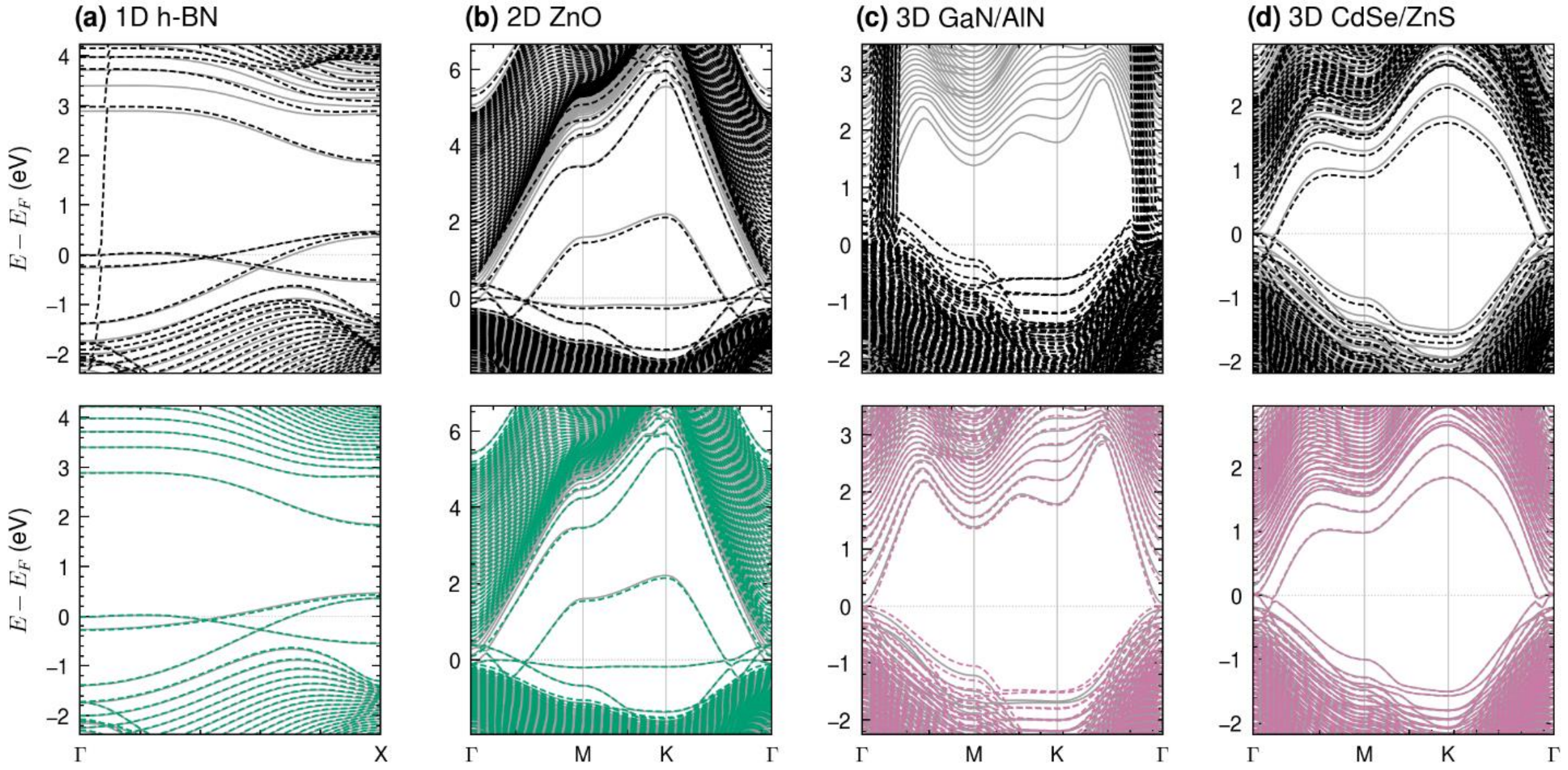

**Figure 4.** Band-structure comparison: (a) 1D h-BN, width 72.3 Å; (b) 2D ZnO, nominal thickness 137 Å; (c) 3D GaN/AlN, $c = 163$ Å and (d) 3D CdSe/ZnS, $c = 159$ Å. DFT bands are gray, and predicted bands are dashed. Each spectrum is referenced to its own Fermi level. Upper panels use the short-range model with three message-passing blocks. Lower panels use the long-range models with the lowest Hamiltonian MAEs bolded in Table 1. The band colors follow the color scheme in Fig. 3.

The improved MLEH-LR Hamiltonians also yield more accurate bands (Fig. 4). For the thickest GaN/AlN superlattice, the TPC lowers the near-Fermi band-energy MAE from 1.37 eV to 88.5 meV. For the h-BN, ZnO, and CdSe/ZnS systems shown in Fig. 4, the band MAEs decrease from 158.2, 170.5, and 107.2 meV to 9.7, 109.6, and 9.6 meV, respectively. Definitions of the Hamiltonian and band errors are given in SI Section S7.

To further examine how the model improves the Hamiltonian and band predictions, we compare the onsite Hamiltonian profiles in Fig. 5, where the long-range models perform much better than the short-range model. The DFT onsite values vary across the polar structures, whereas the short-range predictions show nearly flat interiors or staircase-like transitions. The long-range models reproduce these spatial profiles more closely.

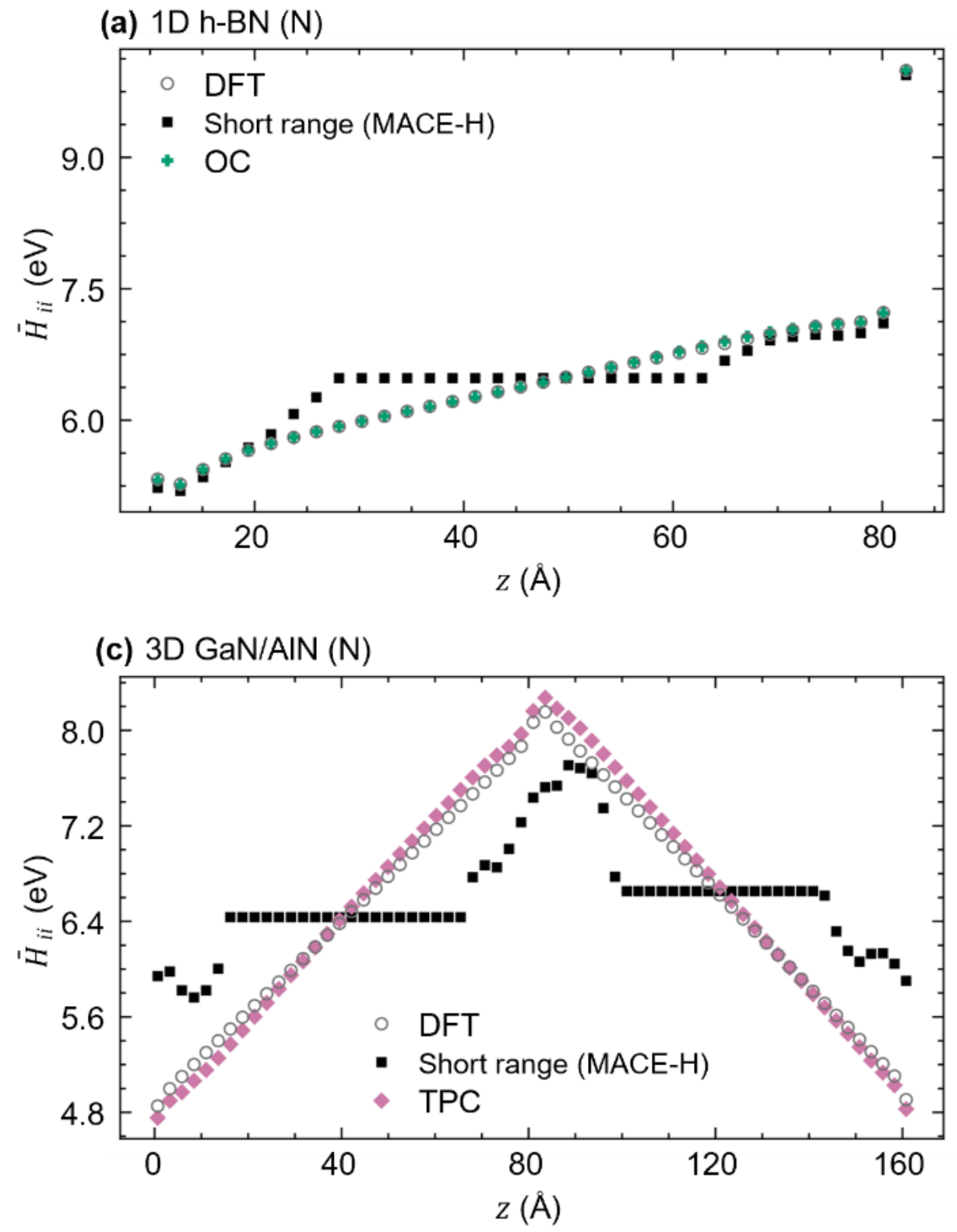

**Figure 5.** Orbital-averaged onsite energies versus position along the long axis. (a) h-BN (width 72.3 Å, N sites), (b) ZnO (nominal thickness 137 Å, O sites), (c) GaN/AlN ($c$ = 163 Å, N sites) and (d) CdSe/ZnS (c = 159 Å, Se sites). Open circles show DFT results, black squares show the short-range model, and diamonds show the long-range model with the lowest Hamiltonian MAE in Table 1 for each system.

### 3.3 Hamiltonians under an applied field

We next examine electronic Hamiltonians under applied electric fields. Previous DFT studies investigated the Stark effect in h-BN ribbons under transverse electric fields $E$ and at different ribbon widths $W$[22–24]. However, the computational cost restricted the widths and fields explored, making dense sampling of the $E$-$W$ space challenging, particularly for large $W$. MLEH-LR enables such high-throughput calculations, allowing us to quantify the joint width and field dependence of the band gap and examine the underlying electronic mechanisms.

Here, we choose H-passivated zigzag h-BN ribbons, which have inequivalent B–H and N–H edges and a built-in transverse potential [Fig. 6(a)]. The finite-field OC model was employed with one message-passing block and one multipole update ($U = 1$). Training covers widths of 79–114 Å and fields $|E| \leq 0.02$ V/Å in either direction. We then test the model at new combinations of widths and field strengths, including a test field magnitude 2.5 times the training limit and a wider 131-Å ribbon, as shown in Fig. 6(b). This model yields Hamiltonian MAEs below 0.76 meV over these test sets. The exact field and width ranges, together with TPC results, are given in SI Section S8.

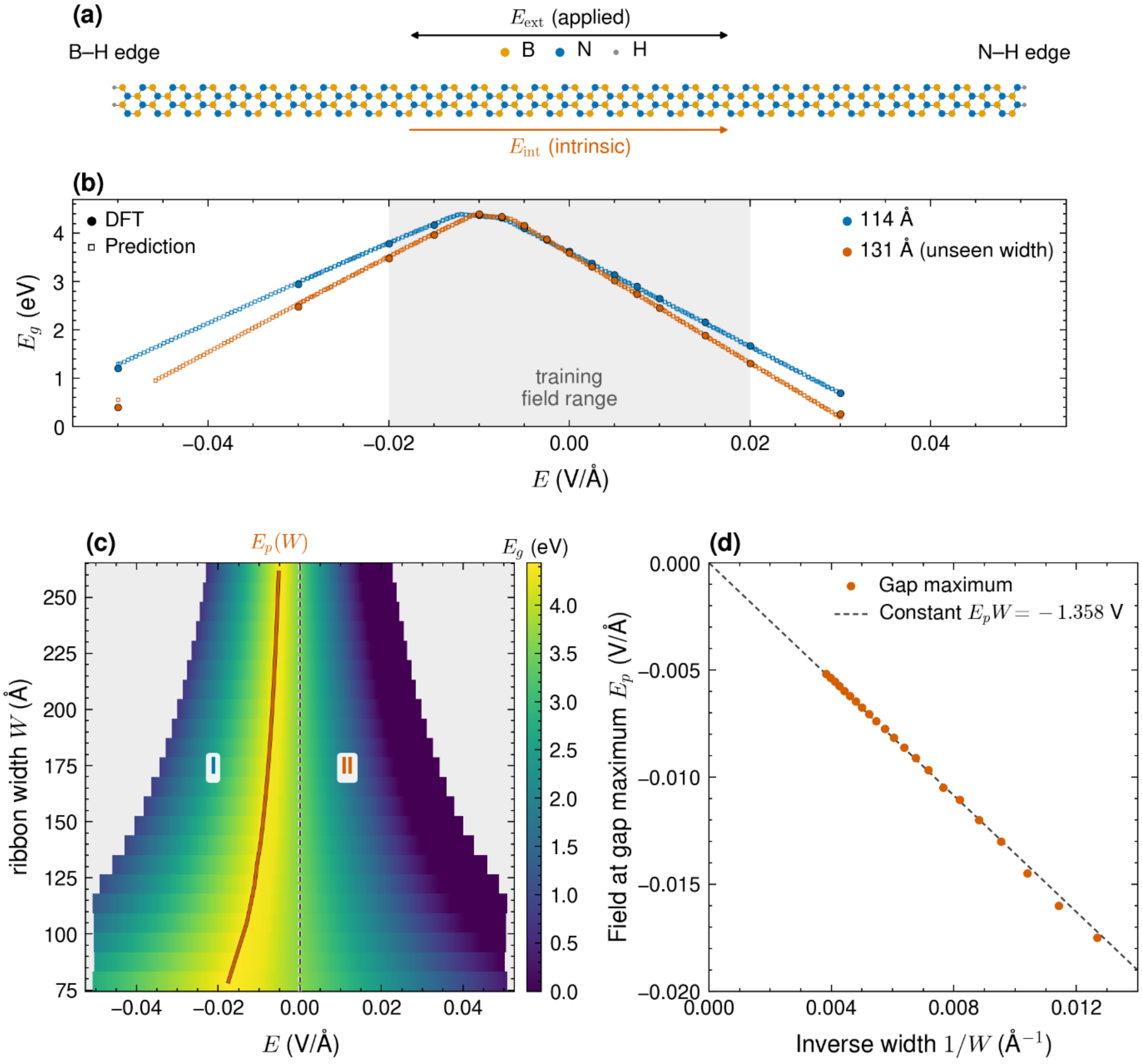

**Figure 6**. External-field benchmark of the OC model on h-BN. (a) Top view of a 131-Å ribbon, in which the positive applied field points in the direction of the indicated intrinsic field. (b) DFT (filled circles) and predicted gaps (open squares) for two widths. Shading marks the training-field range. The 131-Å ribbon tests the model at a width beyond the training range. (c) Predicted gaps over 22 widths. The solid line connects the fields $E_p(W)$ of gap maxima on the computed field grid at each width $W$. Region I, $E < E_p(W)$ and II, $E > 0$ are the regions used to describe the relationship among band gap, width, and electric field in Eq. (11). (d) Field $E_p(W)$ at the gap maximum versus inverse ribbon width $1/W$. Circles show all 22 widths, with $E_p$ determined on the computed field grid. The dashed line shows $E_p = (-1.358\ \text{V})/W$, using the mean $E_pW$ of these points.

We then use this model to calculate the band gap for different combinations of $E$ and $W$ on a dense $E$-$W$ grid containing 5774 points, producing the map in Fig. 6(c). These 5774 Hamiltonian calculations used 774 CPU-core hours, whereas the corresponding DFT calculations would require $2.8 \times 10^4$ CPU-core hours according to our estimate, making MLEH-LR a more efficient approach to high-throughput calculations. With this dense map, we can quantitatively identify which field leads to the largest band gap for each $W$, a result not obtained in previous DFT calculations. Across the 22 widths, we find the relation $E_p W = -1.358 \pm 0.015$ V (Fig. 6(d)), where $E_p$ denotes the electric field leading to the band gap maximum.

To understand the origin of the band-gap maximum, we examine the band structures of a 131-Å ribbon at zero field and three negative fields spanning the gap maximum (Fig. 7). Color represents the average position of each state, indicating its B–H (orange) or N–H (blue) edge character. The transverse field changes the electrostatic potential differently at the two edges, so states concentrated on opposite sides of the ribbon experience different energy shifts. As the field becomes more negative, B–H-edge states are raised relative to N–H-edge states (dashed arrows in Fig. 7(a)). This behavior is clearest in the conduction bands. Initially, the B–H-edge conduction state at X moves upward relative to the VBM, and the gap increases from 3.583 eV at zero field to its maximum sampled value of 4.385 eV at $E = -0.0105$ V/Å. At this field, the B–H edge state at X remains the CBM but is nearly degenerate with the N–H edge state at Q (Fig. 7(c)). At the more negative field of $E = -0.0122$ V/Å, the N–H edge state at Q becomes the CBM, and the gap decreases (Fig. 7(d)). The VBM remains near Q, but its character changes from predominantly N–H-edge-like at zero field to B–H-edge-like near the gap maximum. The gap maximum therefore occurs near the crossover between two competing conduction-band minima: the B–H-edge state at X limits the gap before the crossover, whereas the descending N–H-edge state at Q limits it afterward. Across all tested widths, this crossover occurs near the gap-maximum field characterized by $E_p W = -1.358$ V.

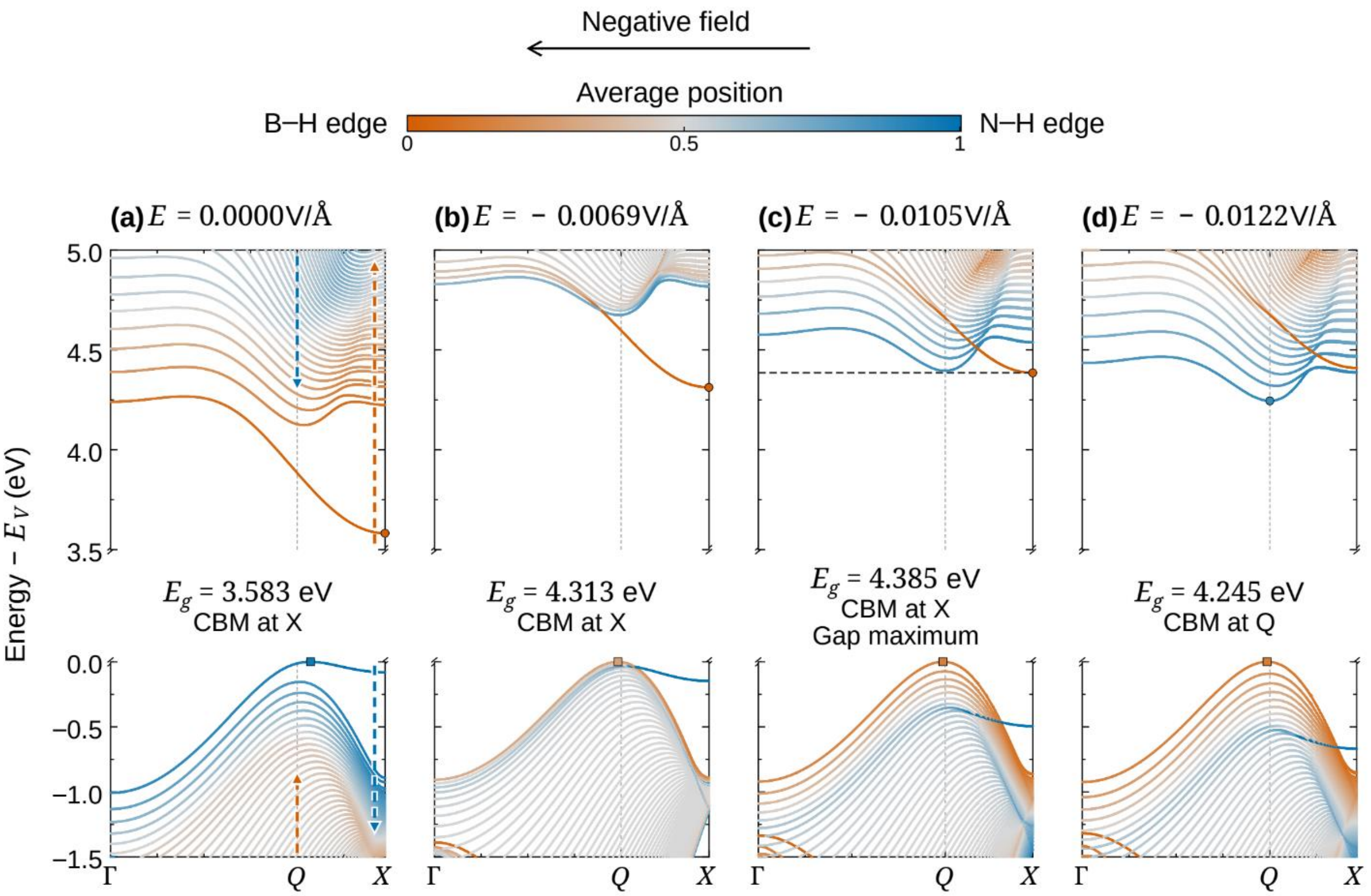

**Figure 7.** Predicted band structures and transverse state distributions of a 131-Å h-BN ribbon at four applied electric fields: (a) $E = 0$, (b) $E = -0.0069$, (c) $E = -0.0105$, and (d) $E = -0.0122$ V/Å. $E_V$ denotes the valence-band maximum (VBM) at each field. Colors encode the normalized position expectation of each eigenstate from the B–H edge (orange) to the N–H edge (blue). Circles and squares mark the conduction-band minimum (CBM) and VBM, respectively. Dashed arrows in (a) illustrate the upward shifts of B–H edge states and downward shifts of N–H edge states as the field becomes more negative. Panel (c) shows the largest gap among the sampled fields, with the CBM at X. The lowest conduction-band energy at Q (4.396 eV) lies close to the CBM energy at X (4.385 eV), marked by the black dashed line.

The dense map of Fig. 6(c) also enables us to formulate a quantitative scaling relation for the band gap dependence on $E$ and $W$. We focus on regions I ($E < E_p(W) < 0$) and II ($E > 0$) in Fig. 6(c), where the band gap changes linearly with E for a fixed W. At E = 0, we expect finite-width effects to diminish as W increases. We therefore propose a 1/W correction to the large-width limits of the extrapolated branch intercepts. At $E \neq 0$, the relative electrostatic shift of spatially separated band-edge states scales with E times their effective separation $(W - \ell_\sigma)$, where $\ell_\sigma$ accounts for the finite spread of band-edge states. These considerations lead us to propose the following expression:

$$E_g^\sigma(E,W) \simeq G_* - \sigma\left(P_\infty - \frac{C}{W}\right) - \kappa e|E|(W - \ell_\sigma). \qquad (11)$$

Here, $e$ is the positive elementary charge, and $\sigma = -1$ and $+1$ in regions I and II, respectively. $G_*$, $P_\infty$, $C$, $\kappa$, $\ell_-$, and $\ell_+$ are the fitting parameters. We fit all six parameters in Eq. (11) jointly to the calculated gap $E_g^\sigma(E,W)$ on the $E$-$W$ grids in regions I and II. Figure 8 shows close agreement between the band gaps calculated with MLEH-LR and those predicted by Eq. (11) across different ribbon widths and electric fields in both regions, demonstrating the accuracy of this relation.

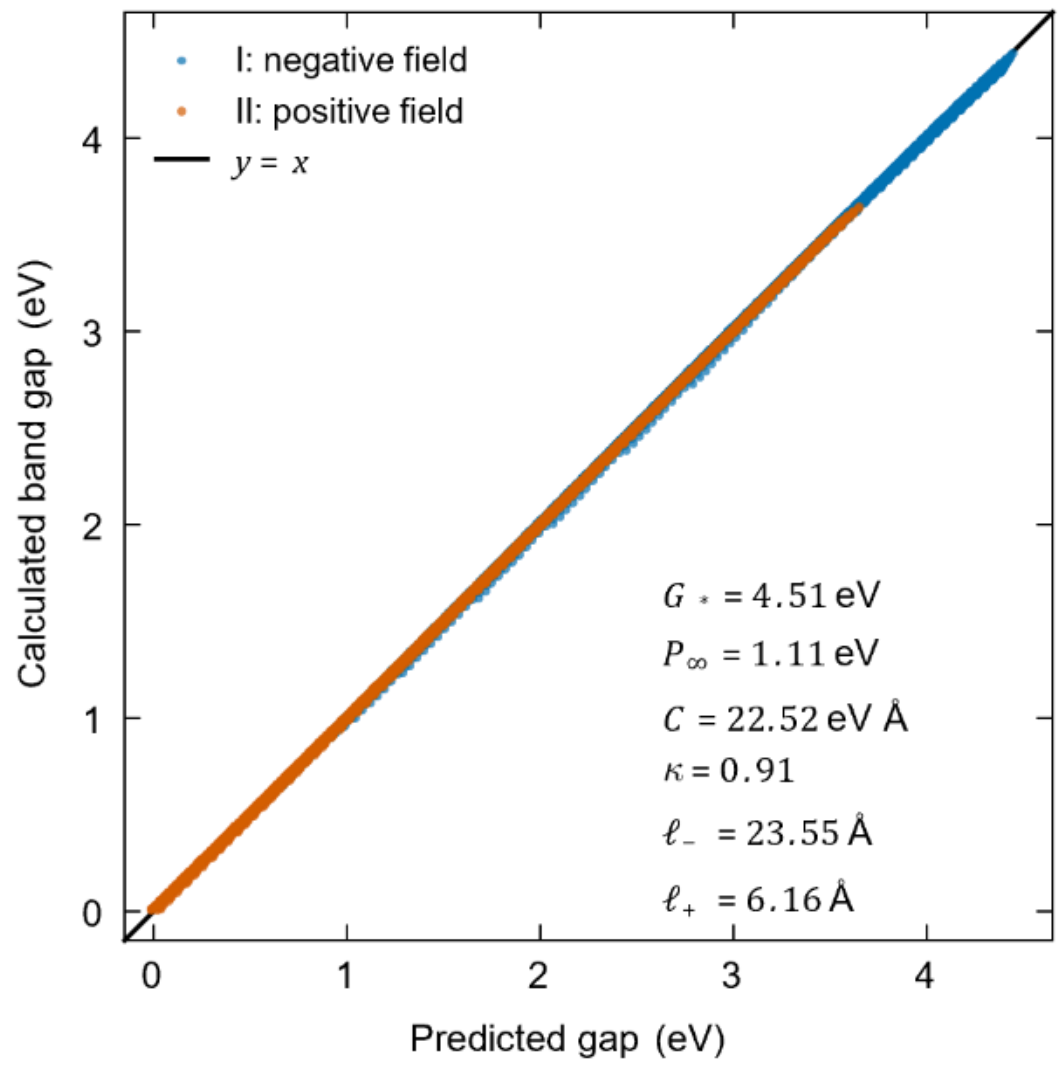

**Figure 8.** Band gaps calculated with MLEH-LR versus those predicted by Eq. (11) for different ribbon widths and electric fields in regions I and II of Fig. 6(c). Blue and orange points denote regions I (negative field) and II (positive field), respectively. The black line indicates perfect agreement. The six fitted parameters are listed in the plot.

## 4 Conclusions

We have developed MLEH-LR to extend machine-learning Hamiltonian prediction beyond local environments by incorporating long-range electrostatics and external fields. Across polar materials spanning one, two, and three dimensions, the framework delivers significant improvements over the short-range model in Hamiltonian accuracy and in reproducing electronic bands.

For materials under electric fields as demonstrated by h-BN ribbons, MLEH-LR enables accurate, large-scale exploration without repeated self-consistent DFT calculations. This exploration reveals the electronic origin of the band-gap maximum and establishes a quantitative relation among band gap, field strength, and ribbon width. These results show how MLEH-LR can uncover physical trends that are difficult to resolve through conventional electronic-structure calculation. More broadly, MLEH-LR provides a route to large-scale, high-throughput studies of polar systems and their response to external fields, enabling accurate and efficient predictions of various properties such as band alignment, interfacial charge redistribution, and carrier transport.

**Acknowledgement**

The code was improved in clarity and consistency with the assistance of Claude and ChatGPT, which also helped in the early stage of manuscript preparation.

**Supporting Information Available**

The Supporting Information is provided with this manuscript. Construction of fixed Gaussian and learned atom-dependent radial functions (Section S1); reciprocal-space Poisson solution and boundary conditions (Section S2); multipole updates with and without external fields (Section S3); implementation details of feature injection (FI), operator correction (OC), and two-stage potential conditioning (TPC) (Section S4); dataset construction, DFT settings, and training parameters (Section S5); FI, OC, and TPC results with the fixed radial function (Section S6); definitions of Hamiltonian and band errors (Section S7); finite-field dataset coverage and TPC results (Section S8); and analysis of the gap maximum across ribbon widths (Figure S1) (PDF).

## Notes

The authors declare no competing financial interest.

## Data Availability Statement

The data supporting the conclusions of this study will be made publicly available after acceptance of the manuscript. The MLEH-LR implementation, the trained model checkpoints, and the training and evaluation configurations required to reproduce the reported results are also available from the corresponding author upon reasonable request.

## TOC Graphic

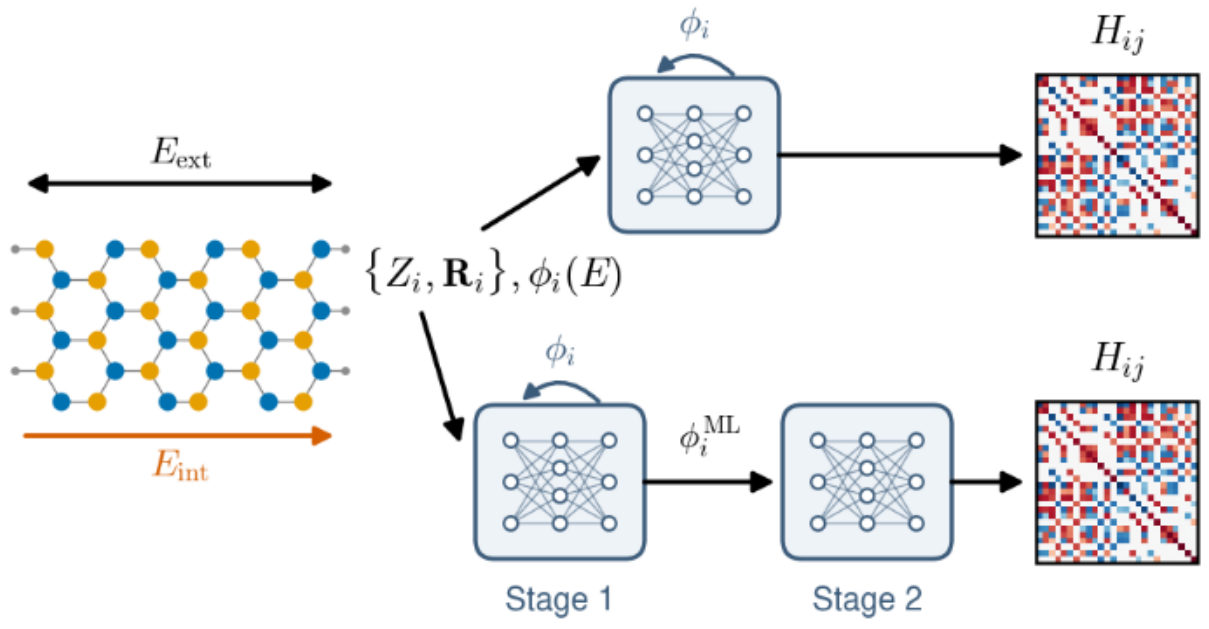